\documentclass[prl,twocolumn,showpacs,superscriptaddress,floatfix]{revtex4-1}

\usepackage{graphicx}% Include figure files
\usepackage{dcolumn}% Align table columns on decimal point
\usepackage{bm}% bold math
\usepackage[usenames]{color}

\begin{document}
\title{High cooperativity coupling of electron-spin ensembles to superconducting cavities}
    \author{D. I. Schuster$^*$}
    \affiliation{Department of Applied Physics and Physics, Yale University}
    \author{A. P. Sears$^*$}    
    \affiliation{Department of Applied Physics and Physics, Yale University}
    \author{E. Ginossar}
    \affiliation{Department of Applied Physics and Physics, Yale University} 
    \author{L. DiCarlo}
    \affiliation{Department of Applied Physics and Physics, Yale University}
    \author{L. Frunzio}
    \affiliation{Department of Applied Physics and Physics, Yale University}
    \author{J. J. L. Morton}
    \affiliation{Department of Materials, University of Oxford, Oxford OX1 3PH, United Kingdom}
    \author{H. Wu}
    \affiliation{Department of Materials, University of Oxford, Oxford OX1 3PH, United Kingdom}
    \author{G. A. D. Briggs}
    \affiliation{Department of Materials, University of Oxford, Oxford OX1 3PH, United Kingdom}
    \author{R. J. Schoelkopf}
    \affiliation{Department of Applied Physics and Physics, Yale University}
\date{\today}

\begin{abstract}
Electron spins in solids are promising candidates for quantum memories for superconducting qubits  because they can have long coherence times, large collective couplings, and many quantum bits can be encoded into the spin-waves of a single ensemble.  We demonstrate the coupling of electron spin ensembles to a superconducting transmission-line resonator at coupling strengths greatly exceeding the cavity decay rate and comparable to spin linewidth.  We also use the enhanced coupling afforded by the small cross-section of the transmission line to perform broadband spectroscopy of ruby at millikelvin temperatures at low powers.  In addition, we observe hyperfine structure in diamond P1 centers and time domain saturation-relaxation of the spins.    
\end{abstract}

\maketitle

An eventual quantum computer, like its classical analog, will make use of a variety of physical systems specialized for different tasks.  Just as a classical computer uses charge-based transistors for fast processing and magnetic hard drives for long term information storage, a quantum computer might use superconducting qubits for processing~\cite{dicarlo_demonstration_2009} and ensembles of electron spins as quantum memories~\cite{imamoglu_cavity_2009,wesenberg_quantum_2009}, linked by single microwave photons.  Although other microscopic systems have been proposed for use in a hybrid architecture~\cite{wallquist_hybrid_2009,taylor_long-lived_2003,verdu_strong_2009,andre_coherent_2006},  electron spins complement superconducting qubits particularly well.  They feature similar transition frequencies, do not require trapping, and can be packed densely.  Furthermore, a single ensemble could be used to store many qubits using holographic encoding techniques~\cite{wesenberg_quantum_2009} demonstrated classically for nuclear~\cite{anderson_spin_1955} and electron~\cite{wu_storage_2009} spins.  

In this Letter, we demonstrate the first step toward realizing a solid-state quantum memory: coupling an electron spin ensemble to an on-chip superconducting cavity at powers corresponding to a single cavity photon.  We observe megahertz spin-photon interaction strengths in both ruby Cr$^{3+}$ spins and N substitution (P1) centers in diamond.  A parallel effort by Kubo, et. al.\cite{XX} sees similar coupling to nitrogen vacancy (NV) centers in diamond.  In doing so we develop a platform for the study of electron spin resonance (ESR) physics in picoliter mode volumes, millikelvin temperatures, and attowatt powers.  Finally, we perform time-resolved saturation/relaxation measurements of the P1 centers, a precursor to full pulsed control of the system.  

\begin{figure}
\includegraphics{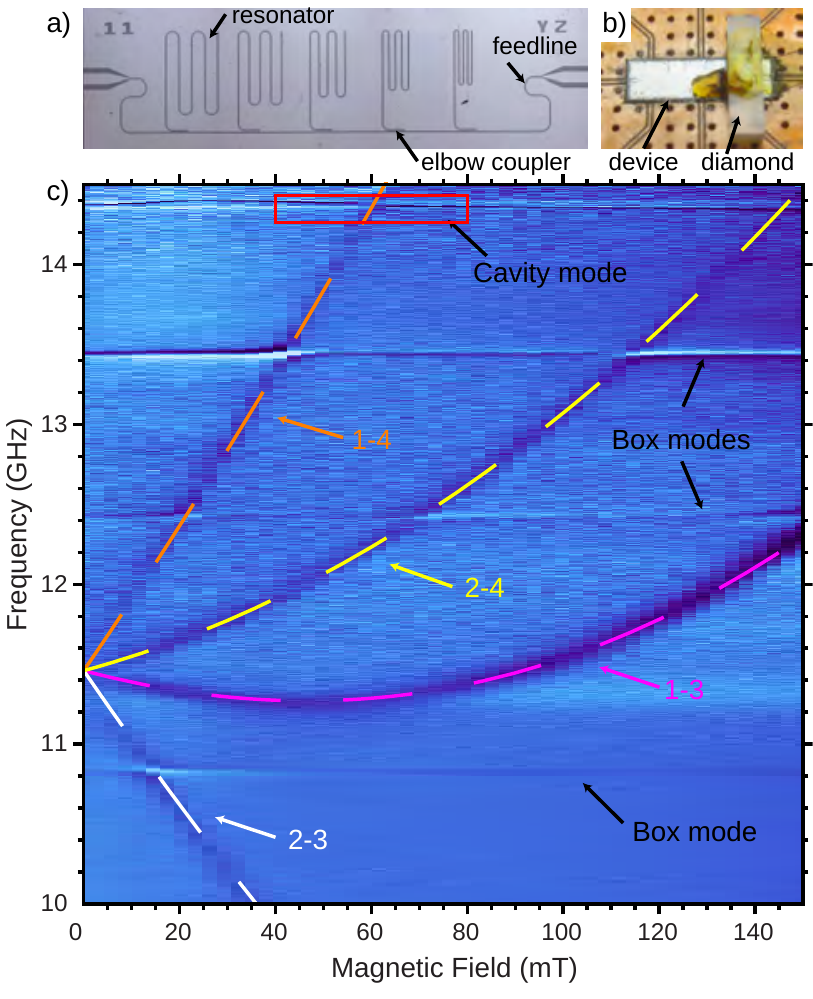}
\caption[Schematic of spin memory device]{Cavity-spin devices and broadband spectroscopy of ruby.   a)  Five  $\lambda/4$ resonators with fundamental frequencies from 4 to 8~GHz coupled to a common feedline fabricated on a ruby substrate.  The coupling Q is determined by the length of the elbow coupler running parallel to the feedline.    b) A piece of synthetic diamond glued on top of a similar device on an undoped c-plane sapphire substrate.  c) The transmission spectrum of the ruby device when a magnetic field is applied parallel to the Nb thin film.  The dashed curves indicate ruby transitions.  Fit uses only two parameters, the magnet current to field ratio and the angle between the magnetic field and crystal $\hat{c}$ axis which is found to be $64.6^\circ$.  To compensate for frequency dependent attenuation in the microwave lines, each point is divided by the mean of all points with the same frequency (row).   The three broad horizontal lines are resonances with the copper sample holder, whereas the narrow line at $\sim 14.35$~GHz is the second mode of the longest superconducting transmission line resonator.  A higher resolution scan of the area within the red box is shown in Fig.~\ref{fig:RubyAvoidedCrossing}a.  }\label{fig:SetupAndBroadband}
\end{figure}

ESR studies the microwave response of electron spins at their resonant frequency in a magnetic field.  Samples are conventionally placed inside a 3D high quality-factor (Q) cavity which enhances the sensitivity by 
confining photons with the spins and extending the interaction time~\cite{Weil2007}.  In this work, several 1D cavities are capacitively coupled to a common feedline on a sapphire chip.  We place the spins within the mode volume by fabricating the device on doped sapphire (ruby -  Fig.~\ref{fig:SetupAndBroadband}a), attaching a substrate on top of an existing device (diamond -  Fig.~\ref{fig:SetupAndBroadband}b), or simply spin-coating the surface (DPPH - not shown).  The single spin-photon coupling is given by $g_{\rm s}/2\pi=m_0 (\mu_0 \omega/2 \hbar V_c)^{1/2}$, where $m_0$ is the spin's magnetic dipole moment.  The single spin coupling can be enhanced by using small mode volume cavities~\cite{boero_electron-spin_2003,narkowicz_scaling_2008}. Using a superconducting cavity allows use of extreme aspect ratios while maintaining high Q.  In this case by using a 1D ($V_c \sim d^2 \lambda$) rather than a 3D ($V_c\sim \lambda^3$) cavity, we increase the coupling by the ratio of the wavelength to the cavity width, $\lambda/d \sim 1000$.   Using an ensemble the effective collective coupling is $g_{\rm{s,eff}}\approx M^{1/2} g_{\rm s} $.  Employing a large number of spins $M\sim 10^{11-13}$ enables megahertz interaction strengths.   Because the resonators are made from superconducting thin films, they can maintain a high Q despite their extreme aspect ratio and $\sim 200$~mT applied in-plane magnetic fields~\cite{mamin_superconducting_2003}.  

Transmission through the ruby chip feedline is plotted in Fig.~\ref{fig:SetupAndBroadband}c from 10 to 14.5 GHz as a function of the in-plane magnetic field. The most striking features are the transitions of the spin-3/2 Cr$^{3+}$ which are visible over a broad frequency range.  

They can be identified by diagonalizing the ruby Hamiltonian~\cite{weber_masers_1959}
\begin{equation}
H_{r}= -m_{0,r} \vec{B}\cdot \vec{S} - D (S_z^2 - \frac{5}{4}), 
\end{equation} 
with $m_{0,r}/2\pi= 27.811 $~MHz/mT, $\vec{S}$ is the spin-3/2 operator, $\hat{z}$ is defined along the $\hat{c}$ axis, and $2 D/2\pi=11.46$~GHz.  In addition to the Zeeman term there is a large crystal field in ruby that separates the levels into two doublets ($S_z=\pm 3/2$ and $S_z=\pm 1/2$).  The measured spectroscopic lines correspond to transitions between the doublets.  When a magnetic field is applied at angle with respect to the $\hat{c}$ axis the levels within each doublet are hybridized giving the observed curvature in the transitions.  Because of the relatively high doping $\sim 10^{19-20}\,  {\rm cm}^{-3}$ and  because the spins efficiently fill the mode volume of the feedline, absorption at the transition frequencies of the Cr$^{3+}$ spins is visible even without the aid of a cavity.  From the $5 \%$  absorption dips we estimate that there are approximately $10^{13}$ spins interacting with the feedline.  

\begin{figure}
\centering
\includegraphics{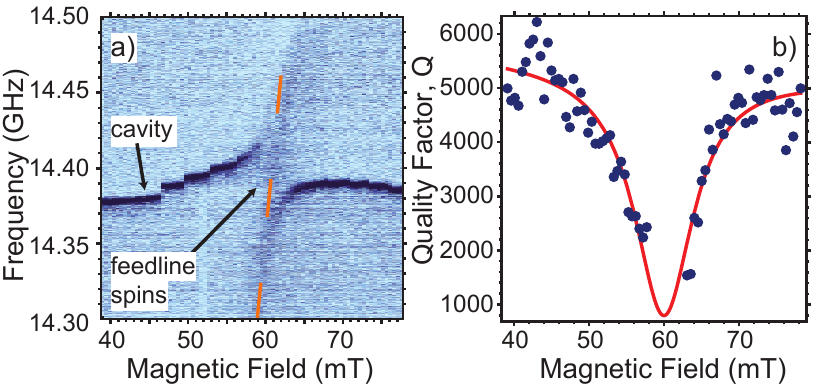}
\caption[Ruby avoided crossing]{a) Transmission spectrum as a function of frequency and magnetic field near resonance with the second mode of one of the superconducting cavities.  The nearly vertical feature in the transmission spectrum at 60 mT is due to the $1\leftrightarrow4$ transition in the spins coupled to the feed line.  An avoided crossing with the superconducting cavity is visible indicating coupling to $M=4\times10^{12}$ spins.  The abrupt jumps in the cavity frequency are thought to be due to flux penetrating the superconducting film causing changes in the effective inductance of the cavity. b) Fit using Eq.~\ref{eq:Q}, to the Q of the cavity, which is partially damped by the spins. }\label{fig:RubyAvoidedCrossing}
\end{figure}

In addition to the spectrum measured via the feedline, several magnetic field insensitive resonances are evident in Fig.~\ref{fig:SetupAndBroadband}c.  The bottom three broad modes are resonances in the sample holder, while the narrow mode at 14.35 GHz is the superconducting transmission line cavity.  Both the modes of the copper sample holder and the superconducting cavities can be used to manipulate spins, but we focus on the on-chip cavities which are higher Q and have typically been used in qubit experiments~\cite{dicarlo_demonstration_2009}.  

In a higher resolution scan (Fig.~\ref{fig:RubyAvoidedCrossing}a) an avoided crossing is present where the $1\leftrightarrow4$ transition approaches the cavity at 14.35 GHz.  Also visible are the interacting spins within the feedline which are unaffected by the cavity-spin coupling.  The cavity line in Fig.~\ref{fig:RubyAvoidedCrossing}a is repelled by more than its linewidth and on resonance it is damped primarily by the spins.  Because of flux jumps (due to a small component of the field perpendicular to the superconducting film), it is difficult to precisely fit frequency versus field.  However, the Q of the cavity is relatively unaffected by these jumps, and can be used to extract the same information.    

The cavity Q is given by
\begin{equation}\label{eq:Q}
Q= \frac{\Delta^2+\gamma_2^2} {2 g_{\rm{s,eff}}^2 \gamma_2 + \kappa \left(\Delta^2+\gamma_2^2\right)}\omega_r,
\end{equation}
where both the spins and the cavity are modeled as single-mode harmonic oscillators with detuning $\Delta$, cavity frequency $\omega_r/2\pi=14.35$~GHz, and spin frequency degenerate at B=60 mT with the spin resonance tuning at rate $m_{\rm eff}/2\pi=52.4$~MHz/mT.  The cavity linewidth $\kappa/2 \pi = \omega_r/2\pi Q = 1.3$~MHz is independently measured away from resonance.  The collective coupling strength and spin decoherence rate are determined from the fit to be $g_{\rm s,eff}/2 \pi=38$~MHz and $\gamma_2/2\pi=96$~MHz.  The latter is probably primarily due to broadening by strong hyperfine interactions with $^{27}$Al nuclear spins~\cite{laurance_aluminum_1962}.  The coupling is much larger than the decoherence rates measured in both the cavity and typical superconducting qubits~\cite{dicarlo_demonstration_2009}.  A dimensionless measure of the coupling strength in cavity QED~\cite{WallsMilburn} is the cooperativity $C=g^2/\kappa \gamma_2 \approx 11.5$.  Since $C>1$ the coupling is strong in the sense that at resonance nearly every photon entering the cavity is coherently transferred into the spins.  In order to make retrieval possible a spin system with smaller linewidth is required.

\begin{figure}
\centering
\includegraphics{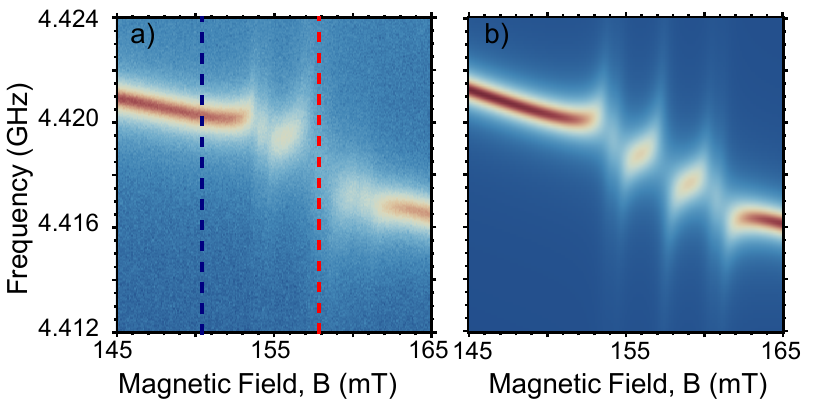}
\caption[Density plots of diamond transmission]{ Experimental (a) and analytically calculated (b) transmission spectrum of cavity-diamond spin system.  In both plots, 5 spin resonances due to the anisotropic hyperfine splitting are observable (see text for details).    The dashed lines in (b) refer to bias fields used in Fig.~\ref{fig:DiamondTimeDomain}.}\label{fig:DiamondDensityPlots}
\end{figure}

Using the diamond sample we are able to observe hyperfine splittings and perform time-domain saturation/relaxation experiments.  ESR active P1 centers~\cite{loubser_electron_1978} are visible via absorption spectroscopy (Fig.~\ref{fig:DiamondDensityPlots}), but we did not observe resonance with NV centers, which have been studied in a parallel publication~\cite{xx}.

The observed splittings can be understood by considering the transitions of the Hamiltonian~\cite{loubser_electron_1978}
\begin{equation}
H_{\rm d}= -m_{0,d} \vec{B}\cdot \vec{S }+ A  \vec{S} \cdot \vec{I},
\end{equation}
where $m_{0,d}/2\pi=28.04$~MHz/mT and the hyperfine coupling tensor $A/2\pi=(81.33, 81.33, 114.03)$~MHz.  Here, the $\hat{z}$ direction corresponds to the diamond 111 axis, $\vec{S}$ are the electron spin-1/2 operators, and $\vec{I}$ are the nuclear spin-1 operators.  This describes a nitrogen atom with an extra, nearly-free electron, and a hyperfine spectrum due to the $I=1$ nuclear spin of $^{14}$N.  This splits the line into three: $m_I=-1,0,1$.  The outer two lines are further split by anisotropy in the hyperfine coupling depending on which carbon the nitrogen substitutional's electron prefers.  Selection rules suppress transitions which change $m_I$, giving a total of five lines shown in  Fig.~\ref{fig:DiamondDensityPlots}b, and observed in Fig.~\ref{fig:DiamondDensityPlots}a.  

The model in Fig.~\ref{fig:DiamondDensityPlots}b treats each of the twelve possible transitions (three hyperfine transitions and anisotropies due to the four bond angles) as an independent harmonic oscillator with each oscillator uniformly coupled to the cavity.  When aligned along one of the bond angles degeneracies reduce the spectrum to the five lines visible in Fig.~\ref{fig:DiamondDensityPlots}.  The effective coupling strengths used were 5 MHz for the central peak, 3.5 MHz for the satellite peaks, along with 30 MHz linewidths for each of the transitions.  The large linewidth is most likely due to additional paramagnetic impurities~\cite{van_wyk_dependences_1997} or magnetic field inhomogeneity.   

\begin{figure}
\centering
\includegraphics{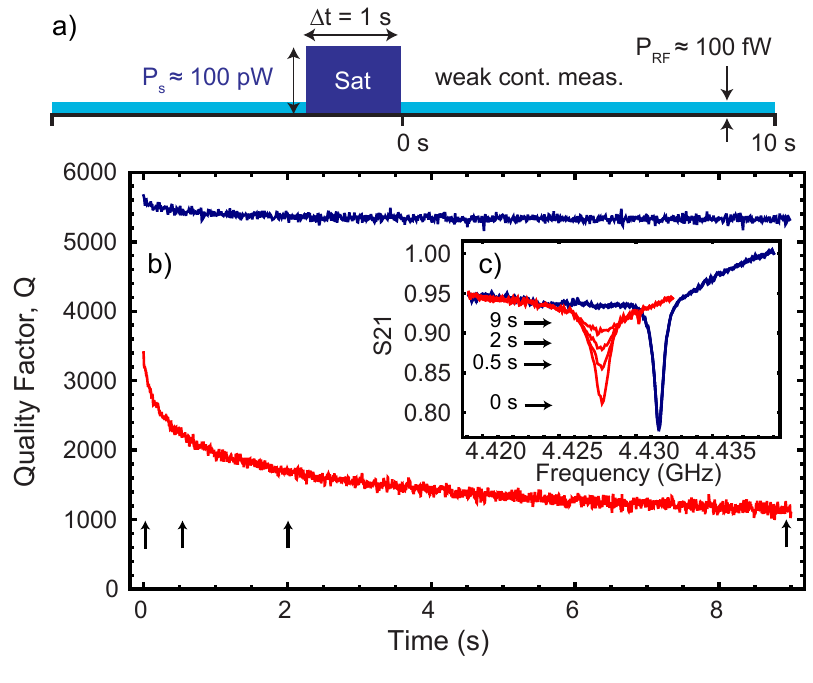}
\caption[Saturation-Relaxation measurement of diamond]{Saturation/relaxation measurement of diamond. a) A saturating pulse ($P\approx100$~pW) is applied to the cavity, depolarizing the spins.  The response of the resonator is probed continuously with a weak probe $P_{\rm RF}\approx 100$~fW which does not significantly excite the spins.   b)~Measurement of resonator Q as a function of time immediately after a 1 second saturation pulse.   The two curves are measured at the locations of the magnetic field indicated in Fig.~\ref{fig:DiamondDensityPlots}b by the dashed lines where the spins are on (red) and off (blue) resonance with the resonator.  c) Measurements were taken at each frequency and the reconstructed cavity resonance fit at each time slice to determine the Q as a function of time.  Each curve corresponds to a different time indicated by an arrow in (b).  The off resonance (blue) curves show almost no time response and lie on top of each other.  }\label{fig:DiamondTimeDomain}
\end{figure}

Although this device was not designed to perform precise rotations on the spins, it is possible to saturate the spins incoherently and observe their recovery in the time-domain.  The data in Fig.~\ref{fig:DiamondDensityPlots}a are measured at 100~fW $\sim 100$~photons in the cavity.  Whereas at low powers the cavity is strongly damped by the spins, saturation begins at $P_{\rm s}=\hbar \omega \gamma \gamma_2 \kappa/2 g_{\rm s}^2\sim10\, {\rm pW}\sim 10^5$~photons and by $P_{\rm s} \approx 100~$pW the cavity Q is restored to nearly its unloaded value.  In Fig.~\ref{fig:DiamondTimeDomain} two experiments are shown at magnetic fields corresponding to the dashed lines in Fig.~\ref{fig:DiamondDensityPlots}b, where the spins are on (red) and off (blue) resonance with the cavity.  This is done at several frequency points near the cavity resonance, allowing us to observe the evolution of the cavity lineshape with time after the saturation pulse.  In Fig.~\ref{fig:DiamondTimeDomain}c one can see that immediately after the saturation pulse the resonator has a high Q.  As the spins relax the resonator becomes strongly damped once more on a timescale of seconds.  A lifetime of $\sim 1$~s is consistent with an estimate based on the saturation power and the linewidth measured in Fig.~\ref{fig:DiamondDensityPlots}~\cite{martinis_decoherence_2005}.   However, it is difficult to extract the relaxation time quantitatively because the curve is not exponential.  This could be attributed to the presence of other ESR centers~\cite{reynhardt_temperature_1998}, heating of the host lattice via the spins, or some other type of saturable impurity which is suspected to play a role in the decay of superconducting cavities~\cite{gao_experimental_2008}.   There appears to be a small time dependence in the off resonance trace in Fig.~\ref{fig:DiamondTimeDomain}b.  This is also consistent with broadband saturable impurities~\cite{gao_experimental_2008,martinis_decoherence_2005} and might point to a technique by which such impurities could be temporarily neutralized.  

Figures~\ref{fig:SetupAndBroadband} - \ref{fig:DiamondTimeDomain} show that these devices can be used as spectrometers of electron spin resonance.  The sensitivity of an ideal ESR spectrometer is a complicated function of both instrument and sample parameters, including ESR linewidths, spin lifetimes (which are themselves temperature and magnetic field dependent).  By substituting in for the mode volume and saturation power the typical sensitivity formula in Weil~\cite{Weil2007} can be re-expressed in the terms of cavity QED, highlighting the fundamental processes in detection:
\begin{equation}
N_{\rm min} = \frac{6 \pi}{s (s+1)}  \left(\frac{g_{\rm s}^2}{\kappa} \tau\right)^{-1/2} \rho_s \left(\frac{2 \gamma_2}{\gamma}\right)^{1/2}  n_a^{1/2}.
\end{equation}
Here, $s$ is the total spin, $\tau$ is the integration time, $\rho_s= \tanh \left(\hbar \omega/k_{\rm B} T_s\right)$ is the polarization of the spins at temperature $T_s$, and $n_a = k_{\rm B} T_N/\hbar \omega$ is the effective number of noise photons added by the detector.  In essence the sensitivity is given by the number of spins required to give $n_a$ scattered photons (at rate $g_{\rm s}^2/\kappa$) in one integration time.  Thermal depolarization ($\rho_s$), and inhomogenous broadening ($2 \gamma_2/\gamma$) worsen sensitivity from the case of ideal spins at zero temperature.   Conventional ESR instruments have $Q = 10,000$, $g_{\rm s}/2\pi \sim 0.02$~Hz ($V_c =10~\mu$L), $T_N = 30000$~K, typical saturation powers of $P_s\sim 100$~mW, and operate from 300~K down to 1~K~\cite{Weil2007}. These typically produce an $N_{\rm min}$ of $\sim 10^9$ (polarized) spins. Our current devices have Q=5000, $g_{\rm s}/2\pi \sim 20$~Hz ($V_c =10~$nL), and $T_N \sim 10$~K, and function from 4~K down to 20~mK. The small mode volume of the 1D resonators also reduces the optimal $P_s$ allowing the study of samples at low temperatures where cooling power is limited. Under similar conditions these parameters predict a spin sensitivity of $N_{\rm min}\sim 10^4$, worsened to about $N_{\rm min}\sim 10^8$ due to the inhomogenous broadening of the spins measured here.  The sample contained approximately $\sim 10^{12}$ which were readily detected even without the use of standard ESR techniques such as field modulation.  

Through further development, future devices could have $Q = 100,000$, $g_{\rm s}=1$~kHz ($V_c = 1$~pL using electron beam lithography), and $n_a=1/2$ using a quantum limited amplifier~\cite{bergeal_phase-preserving_2010}.   For these values $g_{\rm s}/\kappa\sim100$~photons/s, and it should be possible to detect $N_{\rm min} \sim 1$ using a cavity and perhaps even strong coupling if superconducting qubits can be used to mediate the interactions~\cite{twamley_superconducting_2009,marcos_coupling_2010}.

To move forward towards a quantum spin memory several improvements must be made.  Most importantly a more coherent electron spin candidate is required. In addition, because superconducting qubits might be adversely affected by magnetic fields, an ideal candidate would have a ZFS such that at $B=0$, $\hbar \omega \gg k_{\rm B} T$, which could come from crystal field splittings as with the ruby shown here, NV$^-$ centers, etc.  Crystal field split spins have the advantage that they can be tuned with small in-plane magnetic fields allowing in-situ adjustment of the spin frequency and holographic encoding~\cite{wesenberg_quantum_2009}.  Other candidates such as spins with large hyperfine splittings~\cite{wu_storage_2009} might also be interesting as they can be insensitive to small magnetic field fluctuations.  This comes at the expense of not being tunable with such fields.   In addition, they must have long coherence times, requiring control over line broadening due to dipolar couplings, nuclear spins, strain variations (for crystal field splittings) and inhomogeneous bias fields.  New resonator designs could improve the static and microwave magnetic field uniformities, allowing spin-echo techniques to be applied.

We have demonstrated coupling of large ensembles of electron spins to both broadband coplanar waveguide transmission lines and resonators.  The coupling is sufficiently strong to exceed all qubit and cavity decay rates with large cooperativity, but is still limited by the spin linewidth.  The system also shows promise as a general ESR tool allowing good sensitivity in a broadband system, and exquisite sensitivity in a high Q cavity.  The small mode-volume makes it ideally suited for studying picoliter scale samples, especially 2D systems such as graphene or semiconducting heterostructures.  Further applications could include maser amplification and single-photon microwave to optical upconversion 

$^*$D.I.S and A.P.S contributed equally to this work.  The authors would like to acknowledge P. Doering at Apollo Diamond for providing the synthetic diamond sample, G. Ulas for help performing room temperature ESR as well as discussions with S. Lyon, M. Reed, and G. Brudvig.  We acknowledge support for this work is from Yale University (D.I.S.) as well as CNR-Istituto di Cibernetica, Pozzuoli, Italy (L.F.).

%\bibliography{spinspaper}

%merlin.mbs 2010-03-15 4.21a (PWD, AO, DPC)

\end{document}